# Theoretical investigation on isomer formation probability and free energy of small C clusters

Zheng-Zhe Lin*

*School of Physics and Optoelectronic Engineering, Xidian University, Xi'an 710071, China*

*Corresponding Author. E-mail address: linzhengzhe@hotmail.com

**Abstract** – Molecular dynamics simulations and free energy calculations were employed to investigate the evolution, formation probability, detailed balance and isomerization rate of small C cluster isomers at 2500 K. For $C_{10}$, the isomer formation probability predicted by free energy is in good agreement with molecular dynamics simulation. However, for $C_{20}$, $C_{30}$ and $C_{36}$, the formation probability predicted by free energy is not in agreement with molecular dynamics and the detailed balance does not hold, indicating that the molecule system is in non-equilibrium. Such result may be attributed to the transformation barriers between cage, bowl and sheet isomers.

## 1. Introduction

Since decades ago, the structure and preparation of nanoclusters is essential to modern nanotechnology. For example, catalysts used in fuel cells are based on Pt or Pt alloy clusters [1-3], which commonly exhibit a very complex structure. In 1985, $C_{60}$ fullerene was first prepared by Kroto *et al.* using He buffering for small C fragments from laser ablation of graphite. In this famous experiment, the formation probability of $C_N$ clusters was obviously affected by He pressure, and by an integration cup to maximize the cluster reaction, the productions became mainly $C_{60}$ and a few $C_{70}$ [4]. At very high pressure, small C clusters gradually aggregate to large amorphous nanoparticles [5]. Actually, the influence of experimental condition on cluster structure may be difficult to be extrapolated. Therefore, corresponding theoretical investigations should emphasize on the kinetics of formation and isomerization reactions. Slanina *et al.* used isomer free energy to predict isomer formation probability of a given $C_N$ cluster [6]. And such method has been widely applied on various kinds of clusters [7-9]. However, since the topography of global potential energy surface is complicated, the C cluster needs a long time to reach ergodicity [10]. Therefore, the examination of thermal equilibrium and the validity of free energy criteria could be beneficial to corresponding theoretical exploration.

In this work, molecular dynamics (MD) simulation and free energy calculation by Metropolis Monte Carlo method were employed to investigate the evolution, formation probability, detailed balance and isomerization rate of $C_{10}$, $C_{20}$, $C_{30}$ and $C_{36}$ cluster isomers at 2500 K. For $C_{10}$, the isomer formation probability in MD was in good agreement with the theoretical values from free energy calculation. However, for $C_{20}$, $C_{30}$ and $C_{36}$ the free energy criteria failed to predict the formation probability. And the detailed balance between isomers was found not valid, indicating that the system is not in equilibrium. Especially, the cage fullerene of $C_{20}$ [11], which was never found in experimental preparations by laser ablation of graphite, is actually not found in MD. But by free energy criteria its formation probability is not very low. For $C_{36}$, the most probable isomer in MD is the one with D2d symmetry, while the free energy criteria proposes the one with D6h symmetry. These fact prompt us to use the free energy criteria cautiously.

## 2. Methods

*2.1 MD simulation*

To investigate the formation and isomerization of C clusters, MD simulations were performed for isolated C atoms in He buffer gas. The interaction between C atoms was described by the Brenner potential, and Leonard-Jones potential was applied for He-He and C-He interactions. *N* isolated C atoms and 160 He atoms were randomly placed in a cubic with side length of 40 Å (the density of He is about 100 atm at 300 K) and periodical boundary condition applied. Both the C and He atoms were initially set to 2500 K, and a thermal bath at the same temperature was applied on He atoms. By a time step of 0.2 fs, the simulation lasted for 300 ns. During the evolution, the structure of cluster was sampled every 5 ps and cooled down to 0 K. For a given $N$, the formation probability of every isomer was gotten by counting the sampling number in many times of simulations.

## 2.2 *Reaction path and rate*

To obtain the reaction path and corresponding rate of transformation between isomers, MD simulations were performed using the technique in Sec. 2.1, starting from a given $C_N$ cluster isomer instead of C atomic gas, and the time spent on transformation from one isomer to another was recorded. By thousand times of repeated simulations, the probable reaction paths from the given isomer to other isomers were found, and the average reaction rates were derived. Then, the minimum energy paths were calculated by nudged elastic band method [12, 13].

## 2.3 *Free energy*

In equilibrium, the formation probability of cluster isomers corresponds to the free energy. In isothermal-isobaric ensemble, the Gibbs free energy of isomer $a$ and $b$ satisfies

$$G_b - G_a = -kT \ln(N_b / N_a), \quad (1)$$

where $T$ the temperature and $N_a$ and $N_b$ the molecule number of $a$ and $b$. Similarly, in isothermal-isovolumic ensemble the Helmholtz free energy satisfies

$$F_b - F_a = -kT \ln(N_b / N_a). \quad (2)$$

When the clusters are treated as ideal gas,

$$G_b - G_a = F_b - F_a + P_b V_b - P_a V_a = F_b - F_a + kT_b - kT_a = F_b - F_a, \quad (3)$$

and the ratio $N_b/N_a$ in isothermal-isobaric and isothermal-isovolumic ensemble are the same. Then, by $F = -kT \ln Q$, where $Q$ is the molecular partition function, the ratio reads

$$N_b / N_a = e^{-(F_b - F_a)/kT} = Q_b / Q_a. \quad (4)$$

In the following discussion we concern the classical partition function $Q$ to compare with classical MD simulation.

At low temperature, by the rigid-rotor and harmonic-oscillator approximation $Q$ can be decomposed as

$$Q = Q_T Q_R Q_V e^{-E_0/kT}, \tag{5}$$

where $E_0$ the potential energy (PE) of isomer and $Q_T$, $Q_R$ and $Q_V$ the translational, rotational and vibrational partition function, respectively. Here,

$$Q_T = \frac{(2\pi MkT)^{3/2} V}{h^3}, \tag{6}$$

where $M$ the molecular mass and $V$ the volume of simulation box, and

$$Q_R = \frac{\pi^3}{h^3 \delta} \sqrt{(8kT)^3 \pi I_x I_y I_z}, \tag{7}$$

where $I_x$, $I_y$ and $I_z$ the molecular principal moment of inertia and $\delta$ the rotational symmetry number. The quantum mechanical expression for the vibrational partition function reads

$$Q_V = \prod_{i=1}^{3n-6} \frac{e^{-h\nu_i/2kT}}{1-e^{-h\nu_i/kT}}, \tag{8}$$

where $\nu_i$ the canonical vibrational frequency of mode $i$. In the classical limit, it becomes

$$Q_V = \prod_{i=1}^{3n-6} \frac{kT}{h\nu_i}. \tag{9}$$

At high temperature, $Q$ was calculated numerically. For the atoms located at $\vec{r}_1 \sim \vec{r}_n$ with mass $m_1 \sim m_n$ and momentum $\vec{p}_1 \sim \vec{p}_n$, the total energy reads

$$E = \sum_{i=1}^{n} \frac{\vec{p}_i^{\,2}}{2m_i} + U(\vec{r}_1, \vec{r}_2 ... \vec{r}_n), \tag{10}$$

where $U$ the interaction potential, and the classical partition function reads

$$\begin{aligned} Q &= \frac{1}{h^{3n}\delta} \int e^{-E/kT} d\vec{r}_1 d\vec{r}_2 ... d\vec{r}_n d\vec{p}_1 d\vec{p}_2 ... d\vec{p}_n \\ &= \frac{1}{\delta} \left[ \prod_{i=1}^{n} (\frac{\sqrt{2\pi m_i kT}}{h})^3 \right] \left( \int e^{-U/kT} d\vec{r}_1 d\vec{r}_2 ... d\vec{r}_n \right) \end{aligned} \tag{11}$$

Then, to separate the translational motion a new coordinates $\vec{r}_1' = \vec{r}_1$, $\vec{r}_2' = \vec{r}_2 - \vec{r}_1$, $\vec{r}_3' = \vec{r}_3 - \vec{r}_1$ ... $\vec{r}_n' = \vec{r}_n - \vec{r}_1$ are employed. By $U(\vec{r}_1, \vec{r}_2, \vec{r}_3 ... \vec{r}_n) = U(0, \vec{r}_2', \vec{r}_3' ... \vec{r}_n')$ the final factor in Eq. (11) reads

$$\int e^{-U(\vec{r}_1, \vec{r}_2, \vec{r}_3 ... \vec{r}_n)/kT} d\vec{r}_1 d\vec{r}_2 ... d\vec{r}_n$$
$$= \int e^{-U(0, \vec{r}_2', \vec{r}_3' ... \vec{r}_n')/kT} \left| \frac{\partial(\vec{r}_1, \vec{r}_2 ... \vec{r}_n)}{\partial(\vec{r}_1', \vec{r}_2' ... \vec{r}_n')} \right| d\vec{r}_1' d\vec{r}_2' ... d\vec{r}_n', \quad (12)$$
$$= V \int e^{-U(0, \vec{r}_2', \vec{r}_3' ... \vec{r}_n')/kT} d\vec{r}_2' ... d\vec{r}_n'$$

in which the Jacobian $\frac{\partial(\vec{r}_1, \vec{r}_2 ... \vec{r}_n)}{\partial(\vec{r}_1', \vec{r}_2' ... \vec{r}_n')} = 1$. Next, the rotational motion is separated by another transformation stated as follows. From $\vec{r}_2^* = (0, 0, r)$, $\vec{r}_3^* = (\rho, 0, s)$ and arbitrary $\vec{r}_4^* \sim \vec{r}_n^*$, any molecular orientation can be produced by 3-2-3 Euler rotation. Let us rotate $\vec{r}_2^* \sim \vec{r}_n^*$ by $\zeta$ about the z-axis, and by $\theta$ about the y-axis, and then by $\varphi$ about the z-axis. The produced $\vec{r}_i' = R\vec{r}_i^*$ are presented by the rotation matrix

$$R = \begin{pmatrix} \cos\varphi\cos\theta\cos\zeta - \sin\varphi\sin\zeta & -\cos\varphi\cos\theta\sin\zeta - \sin\varphi\cos\zeta & \cos\varphi\sin\theta \\ \sin\varphi\cos\theta\cos\zeta + \cos\varphi\sin\zeta & -\sin\varphi\cos\theta\sin\zeta + \cos\varphi\cos\zeta & \sin\varphi\sin\theta \\ -\sin\theta\cos\zeta & \sin\theta\sin\zeta & \cos\theta \end{pmatrix}. \quad (13)$$

By

$$\vec{r}_2' = R \begin{pmatrix} 0 \\ 0 \\ r \end{pmatrix} = \begin{pmatrix} r\cos\varphi\sin\theta \\ r\sin\varphi\sin\theta \\ r\cos\theta \end{pmatrix} \quad (14)$$

and

$$\vec{r}_3' = R \begin{pmatrix} \rho \\ 0 \\ s \end{pmatrix} = \begin{pmatrix} \rho(\cos\varphi\cos\theta\cos\zeta - \sin\varphi\sin\zeta) + s\cos\varphi\sin\theta \\ \rho(\sin\varphi\cos\theta\cos\zeta + \cos\varphi\sin\zeta) + s\sin\varphi\sin\theta \\ -\rho\sin\theta\cos\zeta + s\cos\theta \end{pmatrix}, \quad (15)$$

the integral element in Eq. (15) reads

$$d\vec{r}_2' d\vec{r}_3' ... d\vec{r}_n'$$
$$= \left| \frac{\partial(\vec{r}_2', \vec{r}_3'... \vec{r}_n')}{\partial(r,\theta,\varphi,\rho,s,\zeta,\vec{r}_4^*,\vec{r}_5^*...\vec{r}_n^*)} \right| drd\theta d\varphi d\rho ds d\zeta d\vec{r}_4^* d\vec{r}_5^* ... d\vec{r}_n^* \qquad (16)$$
$$= \left| \frac{\partial \vec{r}_2'}{\partial(r,\theta,\varphi)} \right| drd\theta d\varphi \cdot \left| \frac{\partial \vec{r}_3'}{\partial(\rho,s,\zeta)} \right| d\rho ds d\zeta \cdot \left| \frac{\partial \vec{r}_4'}{\partial \vec{r}_4^*} \right| \left| \frac{\partial \vec{r}_5'}{\partial \vec{r}_5^*} \right| ... \left| \frac{\partial \vec{r}_n'}{\partial \vec{r}_n^*} \right| d\vec{r}_4^* d\vec{r}_5^* ... d\vec{r}_n^*$$
$$= r^2 \sin\theta drd\theta d\varphi \cdot \rho d\rho ds d\zeta \cdot d\vec{r}_4^* d\vec{r}_5^* ... d\vec{r}_n^*$$

Then, by $U(0,\vec{r}_2',\vec{r}_3'...\vec{r}_n') = U(0,\vec{r}_2^*,\vec{r}_3^*...\vec{r}_n^*)$ the final factor in Eq. (12) becomes

$$\int e^{-U(0,\vec{r}_2',\vec{r}_3'...\vec{r}_n')/kT} d\vec{r}_2'...d\vec{r}_n'$$
$$= \int e^{-U(0,\vec{r}_2^*,\vec{r}_3^*...\vec{r}_n^*)/kT} r^2 \sin\theta drd\theta d\varphi \cdot \rho d\rho ds d\zeta \cdot d\vec{r}_4^* d\vec{r}_5^* ... d\vec{r}_n^* . \qquad (17)$$
$$= 8\pi^2 \int r^2 \rho e^{-U(0,\vec{r}_2^*,\vec{r}_3^*...\vec{r}_n^*)/kT} drd\rho ds d\vec{r}_4^* d\vec{r}_5^* ... d\vec{r}_n^*$$

Combining Eq. (11), (12) and (17) we have

$$Q = \frac{8\pi^2 V}{\delta} \left( \prod_{i=1}^{n} (\frac{\sqrt{2\pi m_i kT}}{h})^3 \right) \left( \int r^2 \rho e^{-U(0,\vec{r}_2^*,\vec{r}_3^*...\vec{r}_n^*)/kT} drd\rho ds d\vec{r}_4^* d\vec{r}_5^* ... d\vec{r}_n^* \right) \qquad (18)$$

and

$$\frac{Q(T_2)}{Q(T_1)} = \frac{\int r^2 \rho e^{-U/kT_2} drd\rho ds d\vec{r}_4^* d\vec{r}_5^* ... d\vec{r}_n^*}{\int r^2 \rho e^{-U/kT_1} drd\rho ds d\vec{r}_4^* d\vec{r}_5^* ... d\vec{r}_n^*}$$
$$= \frac{\int r^2 \rho e^{\frac{U}{k}(\frac{1}{T_1} - \frac{1}{T_2})} e^{-U/kT_1} drd\rho ds d\vec{r}_4^* d\vec{r}_5^* ... d\vec{r}_n^*}{\int r^2 \rho e^{-U/kT_1} drd\rho ds d\vec{r}_4^* d\vec{r}_5^* ... d\vec{r}_n^*}, \qquad (19)$$

whose right-hand side can be treated as the average value of $e^{\frac{U}{k}(\frac{1}{T_1} - \frac{1}{T_2})}$ in the canonical ensemble at $T_1$.

Based on the above, a technique was developed to calculate $Q$ at given temperature. At $T$=300 K, $Q$ was calculated by Eq. (5), (6), (7) and (9). Then, Eq. (19) was employed to precisely calculate $Q$ from low to high temperature. By Metropolis Monte Carlo method, the calculation temperature $T_2$ was increased to 500, 700, 900 ... 2500 K while keeping $T_1=T_2$-200 K, and then the formation probability of every isomer was evaluated by Eq. (4). Note, for isomers with chirality, $Q$ was taken as the sum of both enantiomers.

## 3. Results and discussion

### 3.1 $C_{10}$ cluster

In MD simulation for $N=10$, the C atoms condensed into a $C_{10}$ cluster in about 0.3 ns. Fig. 1(a) shows the evolution of isomer PE sampled in MD, in which the points with a same PE correspond to the same isomer. In most of time, the $C_{10}$ cluster stays at the state with the lowest PE. Sometimes it transforms to a state with high PE and then falls down immediately. By counting the sampling number of each isomer, the 4 isomers of lowest PE are found most probable, denoted as **1**, **2**, **3**, **4** in Fig. 1(b) with their PE and symbols of molecular point group. During the MD simulation, the sum of formation probability for the 4 isomers is about 99.5%, while other isomers with higher PE seldom appear. The relative formation probability of isomer **1**, **2**, **3**, **4** is $0.990 : 5.53 \times 10^{-3} : 1.80 \times 10^{-3} : 2.32 \times 10^{-3}$. And by free energy calculation and Eq. (4), the corresponding theoretical formation probability is $0.991 : 3.01 \times 10^{-3} : 4.57 \times 10^{-3} : 1.60 \times 10^{-3}$, which is in proximity to MD value [Fig. 1(c)] and indicates that the $C_{10}$ system is in thermal equilibrium at 2500 K.

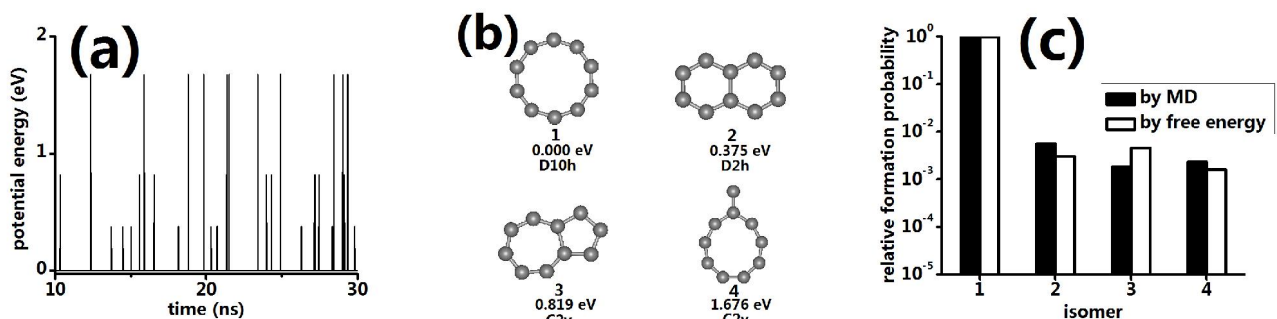

Fig. 1 (a) the PE of $C_{10}$ isomers sampled during the MD simulation, in which the lowest PE is set as 0. (b) the 4 $C_{10}$ isomers of lowest PE, with corresponding PE and symbols of molecular point group marked. (c) the relative formation probability of the isomer **1**, **2**, **3**, **4** in (b) by MD and free energy calculation.

### 3.2 $C_{20}$ cluster

In MD simulation for $N=20$, the C atoms condensed into a $C_{20}$ cluster in less than 0.3 ns, then frequently transformed between the isomers. In the evolution at $T=2500$ K, more than 5000 isomers were found, which can be classified into bowls, sheets and some irregular shapes. By sampling isomer PE, it was found that the isomerization of $C_{20}$ is much more frequently than $C_{10}$, but the $C_{20}$ system seldom stays at the state of the lowest PE [Fig. 2(a)]. The relative formation probability of the 16 isomers of lowest PE appearing in MD is shown by the black columns in Fig. 2(c) (denoted as **1~16**), indicating that the isomer formation probability is obviously unrelated to the level of PE. The structures of isomer **1~7** are shown in Fig. 2(b) with their PE and symbols of molecular point group. By free energy calculation and Eq. (4), the theoretical formation probability of isomer **1~16** (white columns in Fig. 2(c)) is not in accordance with MD values. For some isomers, the difference of formation probability by theory and MD is even in one order of magnitude. It is worth noting that the cage fullerene (denoted as **cage** in Fig. 2(b)), which was considered as the smallest fullerene [11], was not formed in MD like in atomic gas condensation. But the free energy difference between **cage** and **1** is 0.651 eV, corresponding to $N_{\text{cage}}/N_1 = 4.86 \times 10^{-2}$ which is not very low. These fact indicates that the $C_{20}$ system is beyond thermal equilibrium at 2500 K.

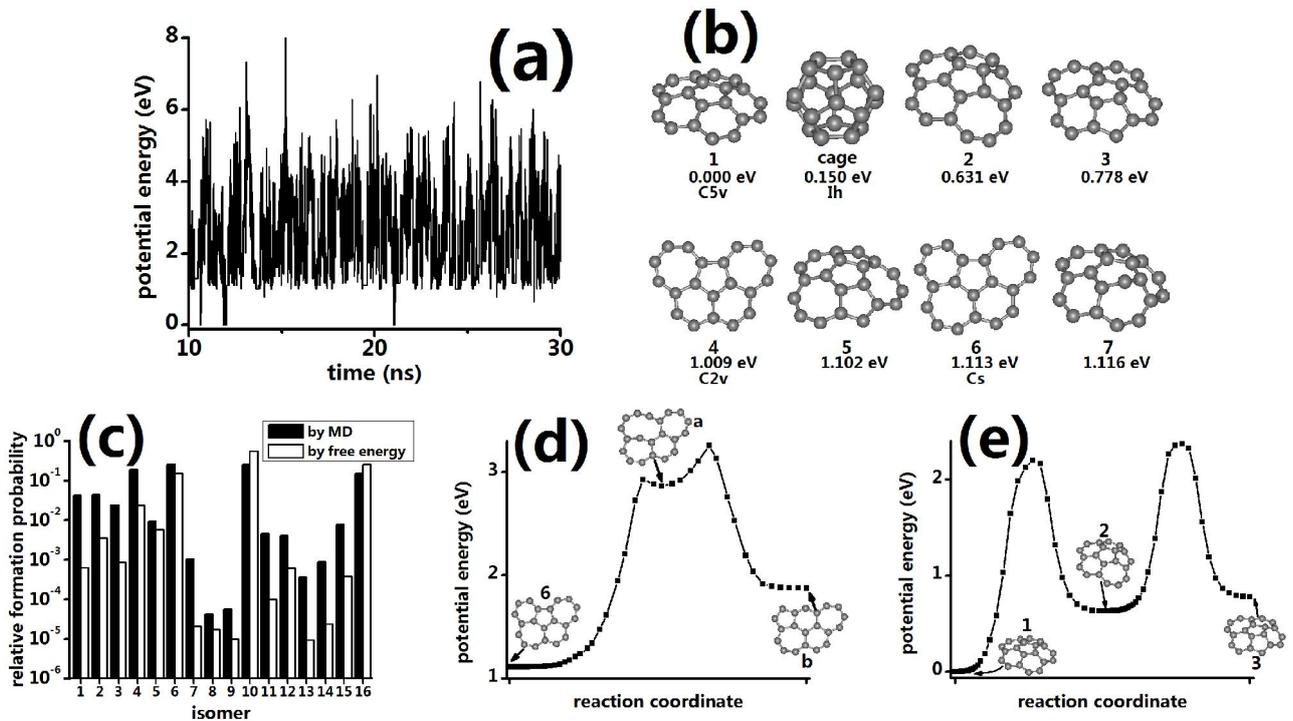

Fig. 2 (a) the PE of $C_{20}$ isomers sampled during the MD simulation, in which the lowest PE is set as 0. (b) the 7 $C_{20}$ isomers of lowest PE (denoted as **1~7**) in MD and the fullerene (denoted as **cage**) of $C_{20}$, with corresponding PE and symbols of molecular point group for symmetric structures marked. (c) the relative formation probability of the 16 $C_{20}$ isomers of lowest PE (denoted as **1~16**) in MD (black column), and corresponding theoretical values by free energy calculation (while column). (d) the potential energy profile of the reaction path from isomer **6** to **b** via an intermediate state **a**, with corresponding isomer structure shown. (e) the potential energy profile of the reaction path from isomer **1** to **3** via an intermediate state **2**, with corresponding isomer structure shown.

The status of $C_{20}$ system away from equilibrium can be further investigated by examining the detailed balance. As an example, we chose the most probable isomer **6** and performed corresponding MD simulation, and a most probable reaction path to an isomer **b** via an intermediate state **a** was found. Fig. 2(d) presents the potential energy profile along the reaction coordinate. In MD simulation, we derived a rate $k_{6\rightarrow a}=4.79\times10^{10}$ s$^{-1}$ for the transformation from isomer **6** to **a**, and $k_{a\rightarrow 6}=2.45\times10^{12}$ s$^{-1}$ for the transformation from isomer **a** to **6**. However, the ratio $N_a/N_6=3.80\times10^{-2}$ of the formation probability of **6** and **a** in MD is away from $k_{6\rightarrow a}/k_{a\rightarrow 6}=1.96\times10^{-2}$ and the detailed balance does not hold. Such situation is also found for the reaction between **a** and **b**, for which $N_b/N_a=2.30$ is also far away from $k_{a\rightarrow b}/k_{b\rightarrow a}=11.7$ ($k_{a\rightarrow b}=2.53\times10^{12}$ s$^{-1}$, $k_{b\rightarrow a}=2.17\times10^{11}$ s$^{-1}$). For the reaction from isomer **1** of the lowest PE to **3** via **2** as an intermediate [Fig. 2(e)], we found $N_2/N_1=1.03$, $k_{1\rightarrow 2}/k_{2\rightarrow 1}=0.278$ and $N_3/N_2=0.558$, $k_{2\rightarrow 3}/k_{3\rightarrow 2}=0.807$. Such deviation between the radio of formation probability and isomerization rate further indicates that the $C_{20}$ system is beyond thermal equilibrium.

3.3 $C_{30}$ *cluster*

In the evolution at 2500 K, more than 9000 isomers were found and they can be also classified into cages, bowls and sheets. Cages has lower PE than bowls and sheets, but the $C_{30}$ system seldom stays at such states. Fig. 3(b) presents the structures of the isomer **a** of the lowest PE and the 6 most probable isomers **b**~**g**. In the upper panel of Fig. 3(a), it can be clear seen that in the evolution the PE of $C_{30}$ is always about 4 eV higher than **a**. The relative formation probability of **a**~**g** in MD is shown by the black columns in Fig. 3(c), which is obviously not in agreement with the theoretical values by free energy calculation and Eq. (4) (white columns in Fig. 3(c)).

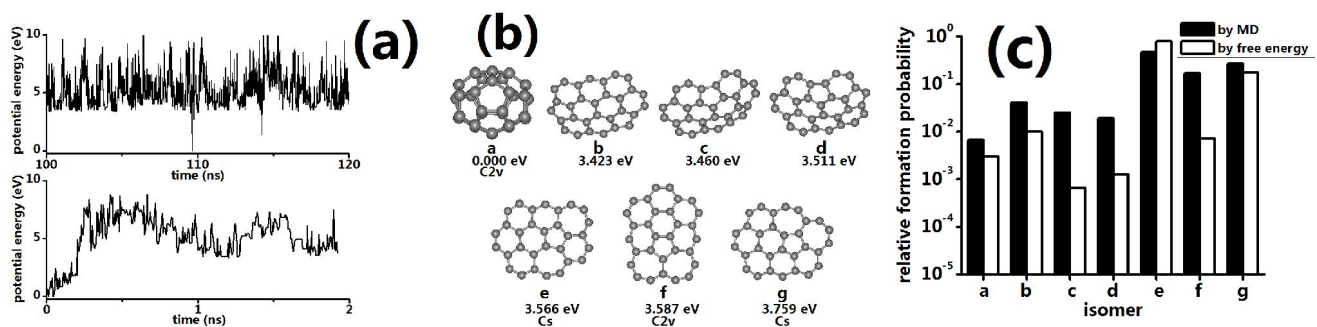

Fig. 3 (a) upper: the PE of $C_{30}$ isomers sampled during the MD simulation. lower: the PE of $C_{30}$ isomers sampled during the MD simulation starting from the isomer of the lowest PE. In both panel the lowest PE is set as 0. (b) the isomers of $C_{30}$: **a** the one of the lowest PE and **b~g** the 6 most probable isomers, with corresponding PE and symbols of molecular point group for symmetric structures marked. (c) the relative formation probability of **a~g** in MD (black column), and corresponding theoretical values by free energy calculation (while column).

We also performed similar MD simulation starting from **a** instead of from atomic gas, and found that **a** quickly transformed into sheet isomer and the PE increased (the lower panel in Fig. 3(a)) in about 2 ns via irregular isomers as intermediate states. And in the following evolution, a spectrum of isomer formation probability similar to previous MD was reproduced. Although free energy of the cage isomer **a** is not the lowest, it is not a stable structure due to some dynamic reason and a non-equilibrium isomer distribution can be always formed.

3.4 $C_{36}$ *cluster*

For $C_{36}$ system at 2500 K, the isomer of the lowest PE, i.e. **1** in Fig. 4(b), can be found in MD [Fig. 4(a)]. The isomer **1** with D6h symmetry is just the one found in experimental preparation [14], but in MD it is less possible than **2** with D2d symmetry. In Fig. 4(b), the 8 most probable $C_{36}$ isomers are shown, in which **1~6** are cages and **7** and **8** are sheets. The corresponding formation probability is shown by the black columns in Fig. 4(c), which is also not in agreement with the theoretical values by free energy calculation and Eq. (4) (white columns in Fig. 4(c)). At 2500 K, the free energy of **1** is the lowest but in MD it is not the most probable one. So, the isomer distribution of $C_{36}$ is also in non-equilibrium.

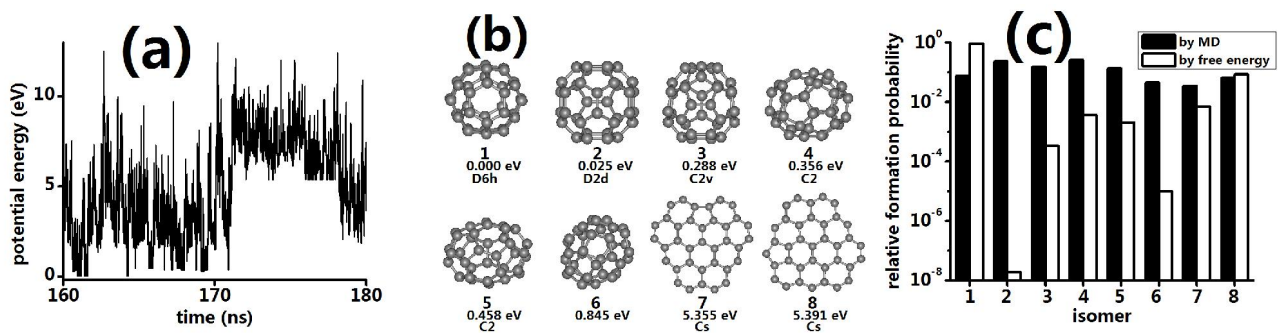

Fig. 4 (a) the PE of $C_{36}$ isomers sampled during the MD simulation. (b) **1~8** the 8 most probable isomers, with corresponding PE and symbols of molecular point group for symmetric structures marked. (c) the relative formation probability of **1~8** in MD (black column), and corresponding theoretical values by free energy calculation (while column).

## 4. Conclusion

In this work, MD simulation and free energy calculation by Monte Carlo method were employed to investigate the evolution, formation probability and isomerization rate of small C clusters at 2500 K. For C clusters of a few atoms, e.g. $C_{10}$, ergodicity is achieved in hundreds of ns and the system is in equilibrium. The isomer formation probability predicted by free energy is in good agreement with MD simulation. However, for larger C clusters, e.g. $C_{20}$, $C_{30}$ or $C_{36}$, the formation probability predicted by free energy is not in good agreement with MD simulation and the detailed balance does not hold, indicating that the thermal equilibrium of the molecule system could not be achieved in several ns. Such result may be attributed to the transformation barriers between cage, bowl and sheet isomers because of their large difference in geometry. The structure transformation from one isomer to another may go through some high-energy intermediate states and thermal equilibrium is difficult to be achieved for so many structures.

***

**Acknowledgements**

This work was supported by the National Natural Science Foundation of China under Grant No. 11304239, and the Fundamental Research Funds for the Central Universities.